\documentclass[lettersize,journal]{IEEEtran}
\usepackage{amsmath,amsfonts}
\usepackage{algorithmic}
\usepackage{algorithm}
\usepackage{array}
\usepackage[caption=false,font=normalsize,labelfont=sf,textfont=sf]{subfig}
\usepackage{textcomp}
\usepackage{hyperref}
\usepackage{stfloats}
\usepackage{url}
\usepackage{verbatim}
\usepackage{graphicx}
\usepackage{booktabs}
\usepackage{cite}
\begin{document}

\title{Augmenting Real-time Visualization in Minimally Invasive Surgery with Particle Filtering and Depth-Aware Occlusion Handling}

\author{Anuska Sarkar, \textit{Graduate Student Member, IEEE,} Partha Acharya, \textit{Member, IEEE,} Vishnu Nandury, Kaustav Basu, and Subhamoy Mandal, \textit{Senior Member, IEEE}
\thanks{All clinical datasets used in this study were collected under the approval of the institutional ethics committee (No. IIT/SRIC/DEAN/2025).}
\thanks{Correspondence to: Subhamoy Mandal (e-mail: smandal@iitkgp.ac.in).}
\thanks{Anuska Sarkar, Partha Acharya and Subhamoy Mandal are with the School of Medical Science and Technology, Indian Institute of Technology Kharagpur, Kharagpur 721302 India (e-mail: asarkar\_24.smst@kgpian.iitkgp.ac.in; Partha.acharya@kgpian.iitkgp.ac.in; smandal@smst.iitkgp.ac.in).}
\thanks{Vishnu Nandury is with the Computer Science Department, Arizona State University, Tempe, Arizona 85287 USA (e-mail: vnandury@asu.edu).}
\thanks{Kaustav Basu is with Narayana Hospital, Kolkata 700107 India (e-mail: dr.kaustavbasu@gmail.com).}}



\maketitle

\begin{abstract}
We present a real-time, hybrid framework for occlusion handling and depth-aware tracking in minimally invasive surgeries (MIS). Occlusions caused by surgical instruments often obscure the region of interest (ROI), affecting surgical precision. To address this, our approach integrates enhanced You Only Look Once (YOLO) based detection, Particle Filtering (PF), and monocular depth estimation using Multiple Depth Estimation Accuracy with Single Network (MiDaS). Occlusion is detected using a distance-based threshold between instruments and ROI. Experimental results demonstrate robust instrument detection with a confidence score of 0.85, effective occlusion event handling, and accurate depth prediction with an MAE of 0.34 cm and RMSE of 0.53 cm. Independent phantom-based validation against stereo depth measurements yielded an MAE of 0.65 cm and RMSE of 0.66 cm. The framework further achieved an occlusion detection accuracy of 81.95\% while maintaining real-time performance at 38.7 FPS. Preliminary surgeon feedback indicates that the proposed visualization framework is clinically relevant and holds promise for future intraoperative image-guided navigation and decision support in MIS.

\end{abstract}

\begin{IEEEkeywords}
Instrument Tracking, Minimally Invasive Surgery, Monocular Depth Estimation, Particle Filtering, Occlusion Handling.
\end{IEEEkeywords}

\section{Introduction}
\IEEEPARstart{M}{inimally} invasive surgeries (MIS) represent a transformative approach to surgical interventions, offering patients a substantial reduction in trauma compared to conventional open surgeries \cite{alfalahi2020concentric}. However, the indirect method of observation and manipulation in MIS complicates the depth perception and impairs the eye-hand coordination of surgeons, which necessitates their acquisition of additional information to monitor the operational instruments moving within the body. Tracking/detection of surgical instruments can provide such important information for operational navigation and visualization in MIS \cite{zhao2017tracking}. In recent years, image-based instrument segmentation and tracking has received increased attention in different minimally invasive scenarios. A recent paper by Rueckert \textit{et al.} provides a comprehensive state-of-the-art review of instrument detection and tracking algorithms, analyzing current approaches for semantic and instance-based segmentation of surgical instruments in endoscopic images and videos while also suggesting potential research directions \cite{rueckert2024methods}.

Particle filtering has been widely utilized across various domains, including human detection and tracking in occluded environments. The basic idea of particle filter is to use a number of independent random variables called particles. In particular, this sequential Monte Carlo approach has been used in parameter estimation and state estimation \cite{chen2003bayesian}. However, its application in surgical tool tracking and visualization remains largely unexplored. Integrating this technique into surgical tool detection using deep learning-based algorithms, thus presents a novel and promising approach. Evidently, in recent years, deep learning-based object detection algorithms like YOLO (You Only Look Once) have been demonstrating exceptional performance in real-time object detection, visualization, and localization with high accuracy \cite{ragab2024comprehensive}. Imaging intelligence powered by AI is revolutionizing medical imaging, enabling enhanced interpretation, real-time analysis, and decision support across the surgical and diagnostic spectrum \cite{mandal2018imaging}. Bodenstedt \textit{et al.} concluded from their validation study that modern deep learning approaches outperform other methods in instrument segmentation tasks, though the results remain imperfect. However, their findings highlight that instrument tracking remains an open challenge, particularly in complex scenarios during conventional laparoscopic surgery \cite{bodenstedt2018comparative}.

Monocular depth estimation in computer vision involves predicting depth information from a single image, rather than relying on multiple viewpoints or stereo vision. This approach enables the reconstruction of 3D spatial structure from 2D videos by estimating relative depth, making it possible to train on diverse datasets without strict calibration requirements or concerns about varying depth ranges across applications. Such methods effectively compensate for the inherent lack of depth perception and visualization in endoscopic and laparoscopic procedures. A robust solution for monocular depth estimation in surgical settings could serve as a crucial foundation for enabling various surgical vision applications \cite{kalim2024imaging}\cite{lach2022local}. Budd \textit{et al.} explored relative monocular depth estimation, which infers depth up to shift and scale from a single image—an active research area. While deep learning models trained on large, diverse meta-datasets achieve excellent performance in natural images, the lack of ground truth depth datasets for endoscopic images makes training such models from scratch impractical. The authors explore the adaptation of general-purpose vision models to the surgical domain and propose an effective method to enhance standard supervision through temporal consistency-based self-supervision \cite{budd2024transferring}.

Despite these advances, instrument occlusion remains a critical issue, potentially obstructing the surgeon’s field of view and limiting effective visualization, thereby leading to procedural errors. Although Jihun Park \textit{et al.} have tackled instrument occlusion by introducing an occlusion-aware loss mechanism, their approach focuses primarily on learning-based strategies \cite{park2024towards}. Our study presents a novel hybrid framework that integrates lightweight particle filtering with instrument detection, monocular depth estimation, and occlusion handling. This synergistic approach improves the robustness of monitoring, visualization, and understanding of a surgical scene, offering a more adaptable and efficient solution for real-time applications.

\section{Methodology}
\subsection{Dataset Description}
We used the publicly available hSDB-Instrument Localization dataset\footnote{\url{https://hsdb-instrument.github.io/}}, which comprises 18,064 RGB images (960 × 540 pixels) extracted from 24 laparoscopic cholecystectomy procedures \cite{yoon2021hsdb}. The dataset captures substantial variability in surgical scenes including different viewpoints, illumination conditions, instrument poses, and tissue appearances encountered during MIS. Each image is annotated with bounding-box labels for 17 surgical instrument classes, including atraumatic grasper (head and body), electric hook, curved atraumatic grasper (head and body), suction irrigation, clip applier (metal; head and body), scissors (head and body), overholt (head and body), ligasure (head and body), clip applier (Hem-O-Lok; head and body), and specimen bag. Since the original hSDB dataset does not include ROI annotations, we manually annotated an additional ROI class using the open-source Computer Vision Annotation Tool (CVAT) to extend it from 17 to 18 classes. The predefined training and validation splits provided with the dataset were used for detector training. To further assess the robustness and generalization capability of the model, an additional five-fold cross-validation study was performed on both the original and the extended datasets. The annotated images were randomly partitioned at the frame level into five folds, with four folds used for training and one fold for validation in each iteration.

For preliminary evaluation, a local video dataset comprising two representative surgical procedures acquired from our clinical collaborator was used exclusively for inference with the best-performing trained model. The first case comprised a robotic-assisted Sentinel Lymph Node Dissection (SLND) procedure\footnote{\href{https://youtu.be/k0359NMMhhM?si=dVihLZFlpDkCbyq0}{Sentinel Lymph Node Dissection (SLND) Robotic Surgery Video}}, performed near the uterine wall, and was selected to demonstrate the proposed framework. To evaluate the applicability of the trained model across a different MIS procedure, an ovarian cystectomy surgical case\footnote{\href{https://youtu.be/-cMkIT2-C00?si=CSkcWWEbgHSZYsUw}{Ovarian Cystectomy Surgical Video}} was also studied which is provided as supplementary material (Fig. S2). All surgeries were conducted under favorable operative conditions, providing clear visualization of anatomical structures and stable camera orientation.

Neither the hSDB dataset nor the clinical surgical videos used for inference provide ground-truth depth annotations. Therefore, relative depth was estimated using MiDaS and calibrated to metric depth, with an independent phantom-based validation performed using a depth camera, as described in Section III(B).

\begin{figure*}[!t]
\centerline{\includegraphics[width=\textwidth]{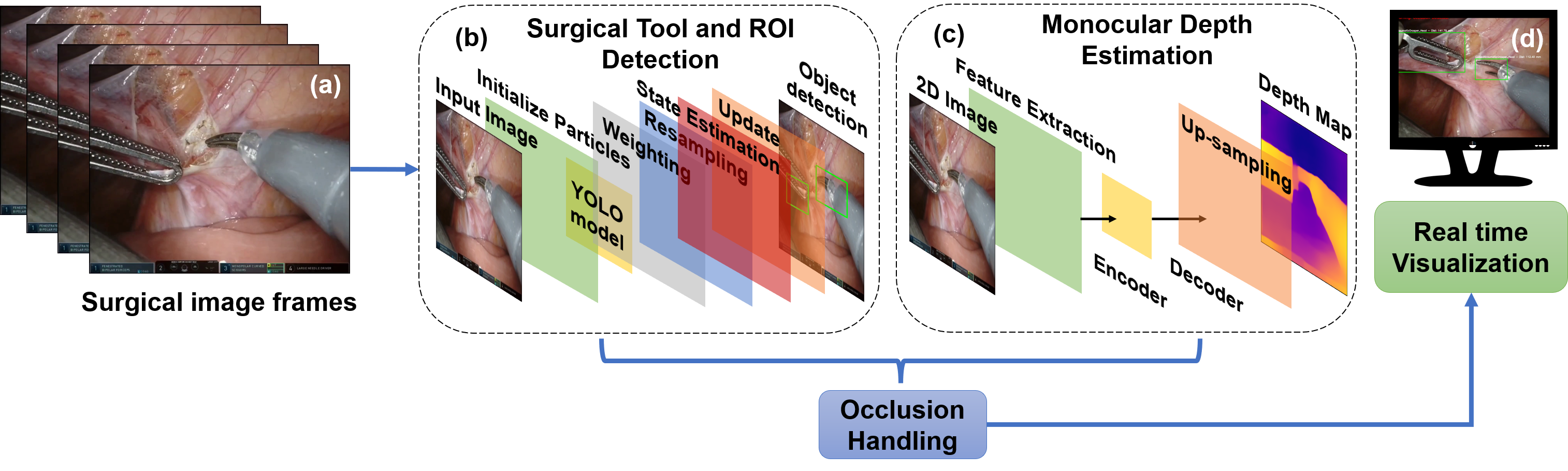}}
\caption{The workflow of the proposed system: (a) Surgical video frames for pre-processing. (b) Integration of particle filtering for surgical tool and ROI detection. (c) Monocular depth estimation using the MiDaS model. (d) Real-time visualization of the surgical scene.}
\label{fig1}
\end{figure*}

Fig. \ref{fig1} illustrates the methodology followed in this study. The process begins with pre-processing, followed by integrating particle filtering with object detection for tracking objects of interest. Simultaneously, monocular depth estimation is applied to provide spatial insights, aiding in occlusion handling. Finally, real-time visualization is improved by integrating detection and depth estimation for better guidance and surgical awareness.

\subsection{Image Pre-processing}
To ensure consistent input formatting, several image pre-processing steps were applied to each ROI within the video frames. Initially, bounding boxes corresponding to detected objects were used to extract cropped segments from the original frame. These cropped regions were then converted from BGR to RGB color space and resized to a fixed resolution using a high-quality Lanczos resampling method. Following resizing, each image was converted into a tensor representation suitable for the model input. Finally, pixel intensity values were normalized and centered around zero for stable inference with the model’s training conditions.

\subsection{Particle Filtering}
Particle filtering (PF) is a computationally lightweight, probabilistic sequential Monte Carlo technique designed for state estimation in nonlinear and non-Gaussian dynamic systems \cite{djuric2003particle}. Instead of propagating a single Gaussian estimate, it represents the posterior probability distribution using a set of weighted particles, each corresponding to a possible system state. This non-parametric representation enables arbitrary posterior distributions to be approximated while simultaneously maintaining multiple competing hypotheses, making PF particularly suitable for object tracking under significant uncertainty and complex motion dynamics \cite{arulampalam2002tutorial,simon2006optimal}.

Surgical instrument tracking during MIS is characterised by highly nonlinear instrument motion, abrupt endoscope movement, temporary occlusions, and non-Gaussian measurement uncertainty caused by smoke, specular reflections, blood, and motion blur. These characteristics motivated the selection of PF over other Bayesian state estimation techniques, such as the Extended Kalman Filter (EKF) and Unscented Kalman Filter (UKF). The EKF estimates the system state by linearising nonlinear process and observation models using first-order Taylor expansion around the current estimate, while propagating a single Gaussian posterior distribution. Consequently, its computational complexity is primarily governed by covariance matrix operations, typically scaling as $\mathcal{O}(n^{3})$, where $n$ denotes the state dimension, and its estimation accuracy may deteriorate under strong nonlinear motion because of linearisation errors \cite{simon2006optimal}. The UKF improves nonlinear estimation by replacing Jacobian-based linearisation with the Unscented Transform, which deterministically propagates $2n+1$ sigma points through the nonlinear system dynamics while retaining a Gaussian posterior representation. Although this generally provides higher estimation accuracy than the EKF for nonlinear systems, its computational complexity also scales approximately as $\mathcal{O}(n^{3})$ owing to repeated covariance propagation and matrix factorisation operations \cite{julier2004unscented}. However, both EKF and UKF inherently assume a unimodal Gaussian posterior and therefore cannot naturally represent the multi-modal probability distributions that frequently arise during prolonged instrument occlusions or intermittent detector failures.

In contrast, PF represents the target state using multiple weighted particles rather than assuming a Gaussian distribution. During temporary detector failures or tissue occlusions, these particles continue to predict the instrument's motion and preserve multiple possible object locations until reliable detections become available again. As a result, it provides continuous tracking in challenging laparoscopic scenes where visual observations are often degraded \cite{arulampalam2002tutorial}. Although PF introduces additional computation due to particle propagation, likelihood evaluation, and resampling, its computational complexity scales approximately as $\mathcal{O}(N_p(C_p+C_w+C_r))$, where $N_p$ denotes the number of particles and $C_p$, $C_w$, and $C_r$ represent the costs of prediction, weight evaluation, and resampling, respectively. Since the proposed framework tracks only a low-dimensional bounding-box state, a particle count of $N_p=100$ was selected empirically because it provided stable tracking while introducing negligible computational overhead. 

The PF algorithm operates through five fundamental steps which collectively enable effective state estimation.
\subsubsection{Creating random particles}Random particles are generated using either a normal distribution (Gaussian distribution) or a uniform distribution to represent the object's state. These particles are based on the belief that they have the potential to represent the object's state. For particles generated using a Gaussian distribution, the process is governed by the mean ($\mu$) and standard deviation ($\sigma$). These statistical measures are computed as follows:

\begin{equation}
    \sigma = \sqrt{\frac{1}{N}{\sum_{i=1}^{N}(x_i - \mu)^2}} 
\end{equation}
\begin{equation}
    \mu=\frac{\sum_{i=1}^{N}{x_i}}{N}
\end{equation}
where \( x_i \) represents each data point and \( N \) is the total number of data points.
\subsubsection{Predict}In the Bayesian framework, the process model is employed to update the belief state of the system. When working with particle filtering, noise is added to the particles after each iteration to account for the inherent uncertainties in the object's motion. This noise models the stochastic nature of the real-world environment, where object movement is rarely ideal. At the end of this step, a set of measurements is obtained, one for each particle distribution. 
The proposal distribution approximates to:
\begin{equation}
    q(x)=p(X_t|X_{t-1},U_{t-1})p(X_{t-1}|d_{0...t-1})d.X_{t-1}
\end{equation}
where  \( q(x) \) is the predictive distribution, \( p(X_t | X_{t-1}, U_{t-1}) \) is the state transition model and \( p(X_{t-1} | d_{0:t-1}) \) is the posterior of \( X_{t-1} \).

\subsubsection{Update}Similar to Bayesian estimation, each particle in particle filtering is associated with a position and a weight. Initially, each particle is assigned an equal weight of 1/N, where \textit{N} represents the total number of particles. Particles with higher weights have a greater likelihood of matching the object, as compared to those with lower weights. During the resampling step, \textit{N} independent random variables are drawn from the particle set, following the conditional distribution defined by their weights.
\begin{equation}
    \sum_{i=1}^{N}\frac{p(y_k|\epsilon_k^j)}{\sum_{j=1}^{N}p(y_k|\epsilon_k^j)}\delta_{s_{k^i}}(dx_k)
\end{equation}
where  \( N \) is the number of particles, \( p(y_k | \epsilon_k^j) \) is the observation likelihood and \( \delta_{s_{k^i}}(dx_k) \) approximates the posterior using weighted particles. 

\subsubsection{State estimation}We track the state of the model after each update with the help of accuracy or error metrics in the same way we estimate the state of the model with the help of mean and variance (Unimodal).

\begin{algorithm}[htb!]
\caption{Particle Filtering for Occlusion Handling}\label{alg:particle_filter}
\begin{algorithmic}
\STATE 
\STATE {\textsc{1. INITIALIZE}}$(N, X_{t-1})$
\STATE \hspace{0.5cm} Set number of particles: \( N \)
\STATE \hspace{0.5cm} Initialize particle positions:
\STATE \hspace{1.0cm} \( X \gets \text{random normal distribution} + \text{initial position} \)
\STATE \hspace{0.5cm} Assign equal weights:
\STATE \hspace{1.0cm} \( w_i \gets \frac{1}{N} \)
\STATE 
\STATE {\textsc{2. PREDICT}}$(X_{t-1}, U_{t-1})$
\STATE \hspace{0.5cm} Apply motion model with noise:
\STATE \hspace{1.0cm} \( X \gets X + U + \text{random noise} \)
\STATE 
\STATE {\textsc{3. UPDATE}}$(y_k, X_t)$
\STATE \hspace{0.5cm} Compute distance to detected position:
\STATE \hspace{1.0cm} \( d_i \gets \|\mathbf{x}_i - \mathbf{y_k}\| \)
\STATE \hspace{0.5cm} Update weights:
\STATE \hspace{1.0cm} \( w_i \gets e^{-d_i} \)
\STATE \hspace{0.5cm} Normalize weights:
\STATE \hspace{1.0cm} \( w_i \gets \frac{w_i}{\sum w_i} \) (if sum is 0, reset to uniform)
\STATE 
\STATE {\textsc{4. RESAMPLE}}$(X_t, w_i)$
\STATE \hspace{0.5cm} Compute effective particle count:
\STATE \hspace{1.0cm} \( N_{eff} \gets \frac{1}{\sum w_i^2} \)
\STATE \hspace{0.5cm} If \( N_{eff} \) is below threshold, resample:
\STATE \hspace{1.0cm} \( X \gets X[\text{resample indices}] \)
\STATE 
\STATE {\textsc{5. ESTIMATE}}$(X_t)$
\STATE \hspace{0.5cm} Compute estimated position:
\STATE \hspace{1.0cm} \( \hat{x} \gets \frac{\sum X}{N} \)
\STATE \hspace{0.5cm} \textbf{return} \( \hat{x} \)
\end{algorithmic}
\end{algorithm}

\subsubsection{Resampling}Initially, all particles have equal weights, but as the algorithm progresses, particles with lower weights are replaced with the ones of higher weights through a resampling process. Common resampling strategies include systematic sampling and stratified sampling, both of which aim to maintain diversity while emphasizing more probable hypotheses. 
Resampling is done when the $N_{\text{eff}}$ value fall below the threshold value and then the resulting distribution is \(P(x)\).
\begin{equation}
    N_{eff} = \frac{1}{\sum_{i=1}^{N} w_i^2}
\end{equation}
where  \( N_{eff} \) is the effective number of particles and \( w_i \) is the weight of the \( i \)-th particle.
\begin{equation}
    P(x_t) = \eta \, p(z_t | x_t) \int p(x_t | u_{t-1}, x_{t-1}) p(x_{t-1} | d_{0:t-1}) \, dx_{t-1}
\end{equation} 
where \( P(x_t) \) is the posterior probability, \(\eta \, p(z_t)\) is the weight and \( p(x_{t-1} | d_{0:t-1}) \) is the prior probability.

\subsection{Surgical Tool \& ROI Detection}
The tool and ROI detection process begins by passing an input frame to the YOLOv10 model for prediction. The model generates detection results, including bounding boxes, and class predictions for each instrument and ROI present in the frame. For each detections, the bounding box coordinates \((x_1,y_1,x_2,y_2)\) are extracted in its prediction. The dimensions of the bounding box are calculated by determining its width \(w\) and height \(h\), where \(w=x_2-x_1\) and \(h=y_2-y_1\). The final output contains the class ID, class name, and bounding box of the detected objects.
The bounding boxes serve as initialization points for PF, which then estimates the object's state over consecutive frames.

\begin{figure*}[!t]
\centerline{\includegraphics[width=\textwidth]{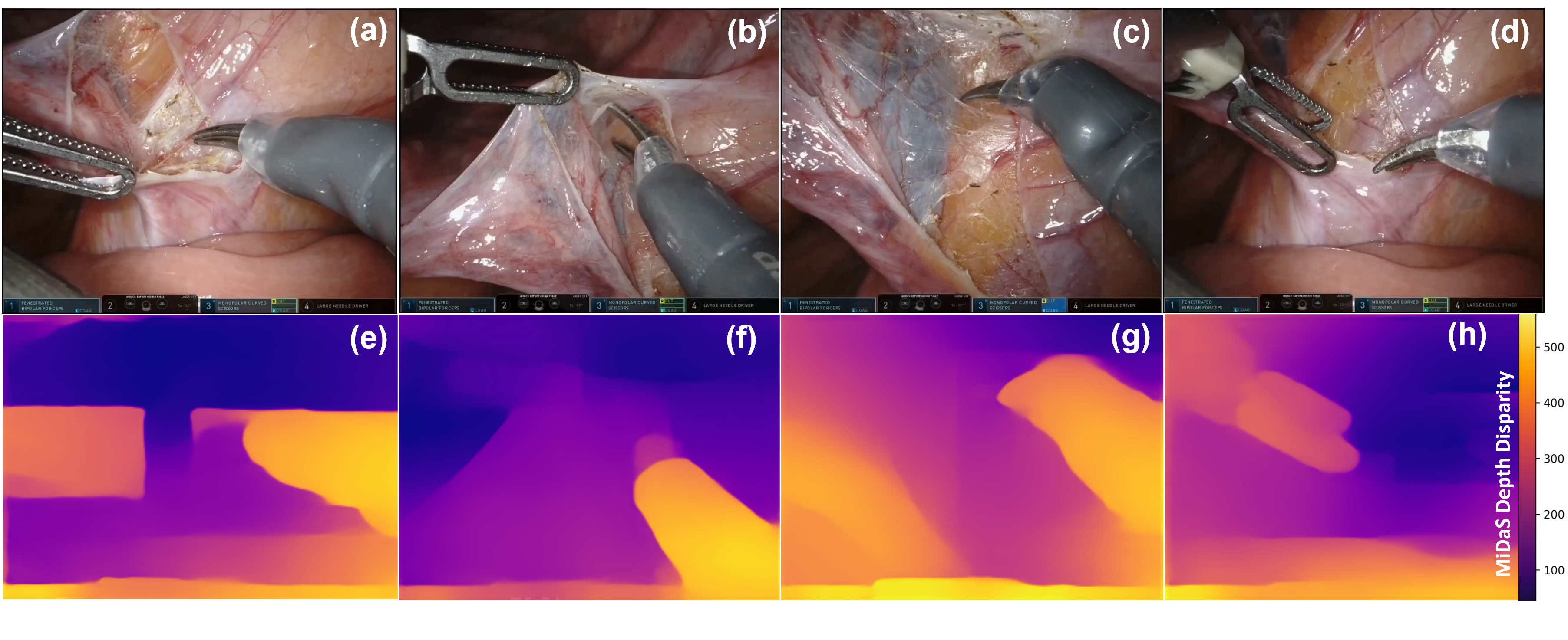}}
\caption{Representative frames of different surgical actions acquired from robot-assisted cholecystectomies (a–d) with corresponding relative depth maps (e–h).}
\label{fig2}
\end{figure*}

\subsection{Monocular Depth Estimation} In the context of MIS, visual occlusions occur when a surgical instrument obstructs the line of sight to a tissue landmark or ROI. While traditional 2D object detection identifies the spatial location of these elements within a frame, it cannot provide information regarding their distance from the laparoscopic camera. To overcome this limitation, monocular depth estimation is employed to recover the relative scene geometry from a single RGB image. The resulting depth map is subsequently calibrated to approximate real-world depth measurements, thereby providing quantitative spatial information for each detected instrument and ROI. These calibrated depth estimates enhance the interpretation of the surgical scene by conveying the relative positioning of anatomical structures and surgical instruments during navigation.

In our approach to depth estimation, we have leveraged the MiDaS (Multiple Depth Estimation Accuracy with Single Network) model for monocular depth estimation, which provides pixel-wise relative depth information from a single RGB image \cite{budd2024transferring}\cite{fredin2024estimating}. The model is pre-trained on diverse datasets and estimates the relative spatial relationship of each pixel with respect to the camera, producing a dense depth representation of the scene. The predicted values are dimensionless and preserve the ordinal depth relationships among scene points, where higher values correspond to surfaces that are relatively closer to the camera. Consequently, the generated depth map captures the 3D structure of the surgical scene but does not directly provide metric depth measurements. Metric depth is subsequently obtained through the calibration procedure described in Section III-B. Fig. \ref{fig2} presents example images from the laparoscopic domain, illustrating both the original surgical frames and their corresponding relative depth representations. Additionally, a supplementary set of example images is provided (refer to Fig. s1) to further illustrate the relative depth representations.
\par Given a bounding box defined by its corners \((x_1,y_1)\) and \((x_2,y_2)\), are clipped to ensure they lie within the image frame dimensions such that \(x'_1=max(0,x_1)\), \(y'_1=max(0,y_1)\) and \(x'_2=min(width,x_2)\), \(y'_2=min(height,y_2)\) where width and height are the dimensions of each frame. Once the ROI is extracted from the cropped image, the model is integrated to generate a depth map for each frame. The depth map values are normalized to the range [0,1]: 
\begin{equation}
    D_n(x,y)= \frac{depthMap(x,y)-min(depthMap)}{max(depthMap)-min(depthMap)}
\end{equation}
where  \(D_n(x,y)\) is the normalized depth value and \((x,y)\) are pixel coordinates in the depth map.
The normalized depth values are further scaled to a predefined range in millimeters (mm) such that:
\begin{equation}
    D_s(x,y)=D_n(x,y)\times S
\end{equation}
where \(D_s(x,y)\) is the scaled depth value and \(S\) is the scaling factor (in mm).

The final depth estimate is computed as the average depth within the bounding box over an area \(A\): 
\begin{equation}
    D=\frac{1}{A}\sum_{(x,y)} D_s(x,y)
\end{equation}

The calibrated depth associated with each detected instrument and ROI is subsequently displayed in the augmented visualization and recorded for quantitative analysis throughout the surgical sequence.

\subsection{Occlusion Handling} Occlusion detection is determined from the image-plane proximity between the PF estimated positions of the instrument and the ROI. The resulting occlusion status is complemented by calibrated monocular depth estimates to provide additional spatial context during visualization. PF estimates the object states using a set of particles, each representing a potential state hypothesis. At each time step, the filter predicts the state based on the previous estimate and the motion model, and subsequently updates the particle weights according to the likelihood of the current observation. For the instrument, the predicted state at time \(t\) is:
\begin{equation}
    \hat{x}_i(t)=\sum_{i=1}^{N}w_{i}^t\epsilon_i(t)
\end{equation}
where \(\hat{x}_i(t)\) is the estimated state of the instrument at time \(t\), \(\epsilon_i(t)\) is the state of the \(i^{th}\) particle of the instrument and \(w_{i}^t\) is the weight associated with the \(i^{th}\) particle.
Similarly, the predicted state of the ROI at time \(t\) is given as \(\hat{x}_r(t)\).

Once the positions of both the instrument and the ROI are estimated by the particle filter, the distance \(d(t)\) between the instrument and the ROI is computed using the Euclidean distance between their estimated states: 
\begin{equation}
    d(t)=||\hat{x}_i(t)-\hat{x}_r(t)||
\end{equation}
Occlusion is detected when the distance \(d(t)\) between the instrument and the ROI is less than a predefined threshold \(\delta\). 
The occlusion detection criterion is expressed as:
\begin{equation}
\text{occlusion}(t) =
\begin{cases} 
\text{True}, & \text{if } d(t) < \delta \\
\text{False}, & \text{if } d(t) \geq \delta
\end{cases}
\end{equation}
The system ensures that if an occlusion is detected, an alert is displayed in the output video, which helps the surgeon or system to take preventive or corrective measures.

\section{Experiments}
\subsection{Implementation Details}
The experiments were implemented using the PyTorch framework on a local workstation equipped with an NVIDIA GeForce RTX 4060 Laptop GPU with CUDA acceleration. The detector was trained using the YOLOv10 architecture with the AdamW optimizer, an initial learning rate of 0.001, and a total of 100 training epochs. The model was trained on images resized to 960 × 960 pixels with a batch size of 8, a patience value of 5 for early stopping, and Automatic Mixed Precision (AMP) enabled to improve computational efficiency. Notably, no data augmentation was applied to preserve anatomical accuracy and clinical realism, as artificial transformations could introduce unrealistic cues that may distort depth estimation and spatial understanding.

The best-performing model obtained from training on the hSDB dataset was subsequently deployed on the local surgical dataset for inference without further fine-tuning to assess the evaluation of the proposed framework on unseen clinical videos. During inference, the detector outputs were temporally refined through particle filtering, followed by occlusion analysis to identify instrument-ROI interactions and generate real-time occlusion alerts. Relative depth information was estimated using the MiDaS model and integrated with the detection pipeline to support depth visualization and occlusion guidance. The implementation was designed as a modular processing pipeline, allowing detection, tracking, occlusion analysis, and depth estimation to operate seamlessly in real time.
\begin{figure*}[!t]
\centerline{\includegraphics[width=\textwidth]{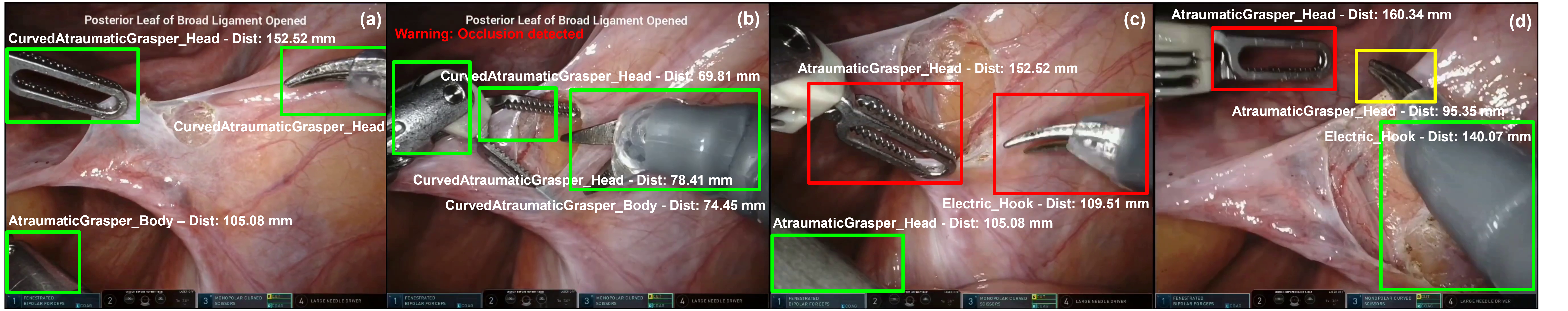}}
\caption{Representative surgical frames illustrating real-time instrument tracking and occlusion handling at (a) t, (b) t+3s with a text-based alert, (c) t+5s with a color-coded bounding box, and (d) t+9s with depth-aware color-coded cues, where the bounding box changes from green (safe) to yellow (partial occlusion) and red (complete occlusion).}
\label{fig3}
\end{figure*}

\begin{table}[!t]
\caption{Class-wise detection performance of the proposed YOLOv10 detector on the extended hSDB dataset.}
\label{tab1}
\centering
\scriptsize
\setlength{\tabcolsep}{5pt}
\renewcommand{\arraystretch}{1.05}
\begin{tabular}{lcccc}
\hline
\textbf{Class} & \textbf{Precision} & \textbf{Recall} & \textbf{mAP$_{50}$} & \textbf{mAP$_{50-95}$} \\
\hline
\textbf{Overall} & \textbf{0.793} & \textbf{0.867} & \textbf{0.894} & \textbf{0.713} \\
\hline
ROI & 0.505 & 0.319 & 0.356 & 0.102 \\
AtraumaticGrasper\_Head & 0.894 & 0.913 & 0.959 & 0.760 \\
AtraumaticGrasper\_Body & 0.858 & 0.900 & 0.949 & 0.757 \\
Electrichook & 0.925 & 0.971 & 0.984 & 0.862 \\
CurvedAtraumaticGrasper\_Head & 0.808 & 0.898 & 0.908 & 0.718 \\
CurvedAtraumaticGrasper\_Body & 0.753 & 0.932 & 0.948 & 0.774 \\
Suction-Irrigation & 0.896 & 0.940 & 0.944 & 0.826 \\
ClipApplier(Metal)\_Head & 0.944 & 0.738 & 0.863 & 0.582 \\
ClipApplier(Metal)\_Body & 0.881 & 0.947 & 0.940 & 0.829 \\
Scissors\_Head & 0.845 & 0.776 & 0.873 & 0.599 \\
Scissors\_Body & 0.934 & 0.857 & 0.931 & 0.764 \\
Overholt\_Head & 0.695 & 0.944 & 0.952 & 0.648 \\
Overholt\_Body & 0.654 & 1.000 & 0.906 & 0.808 \\
Ligasure\_Head & 0.945 & 0.934 & 0.981 & 0.782 \\
Ligasure\_Body & 0.971 & 0.967 & 0.986 & 0.915 \\
ClipApplier(Hem-O-Lok)\_Head & 0.865 & 1.000 & 0.995 & 0.699 \\
ClipApplier(Hem-O-Lok)\_Body & 0.258 & 1.000 & 0.995 & 0.995 \\
SpecimenBag & 0.640 & 0.561 & 0.615 & 0.413 \\
\hline
\end{tabular}
\end{table}


\subsection{Monocular Depth Calibration}
\subsubsection{Phantom-based Calibration}
To validate the calibrated metric depth estimates, an independent phantom-based experimental setup was developed, as shown in Fig. \ref{fig8}. A standard laparoscopic box trainer was used to simulate the MIS workspace, within which a tissue phantom was positioned. Reference depth measurements were acquired using an OAK-D Lite stereo RGB-D camera (Luxonis, Boulder, CO, USA), comprising a 12 MP RGB sensor and a pair of monochrome stereo cameras. The camera was rigidly mounted above the phantom at a representative laparoscopic working distance to emulate the viewpoint encountered during MIS. An ArUco fiducial marker was placed at multiple locations on the phantom surface to provide reproducible reference points for depth measurement. For each marker position, paired RGB images and corresponding stereo depth maps were acquired. The RGB images were processed using the calibrated MiDaS model to estimate metric depth, while the corresponding stereo depth maps served as the independent reference measurements. The ArUco marker was automatically detected in each image, and the median depth within the marker region was extracted from the stereo depth map and compared with the corresponding calibrated MiDaS metric depth estimate.

\subsubsection{Clinical Validation}
To obtain depth information from surgical scenes, we employed the MiDaS model for monocular depth estimation. It provides relative depth disparity maps, which were subsequently calibrated to real-world measurements using a manual scaling process. A sterile paper scale was introduced into the surgical field, attached to an instrument, and served as a physical reference. Known coordinate points were manually marked on this scale, allowing us to map MiDaS-derived depth values to corresponding real-world values, as shown in Fig. \ref{fig5}.
Since MiDaS outputs only relative depth, we applied min-max normalization to transform the predictions into absolute depth values within the expected surgical depth range. The normalized values were scaled to match an empirically chosen range of 1 cm to 6.5 cm which served as an approximation of the typical distance between the laparoscopic camera and the region of interest (ROI). The distance was estimated by visually inspecting surgical frames while varying the surgical position and camera scope orientation.

\section{Results}
\subsection{Detector Performance}
Table. \ref{tab1} presents the performance metrics of the trained model in hSDB dataset for detecting various surgical instruments and ROI. Metrics such as precision and recall assess the model’s accuracy in identifying true positives while minimizing false positives and negatives. The mean Average Precision at IoU 0.5 (mAP50) indicates how well the model detects objects with at least 50\% overlap with ground truth, whereas mAP50-95 averages precision across multiple IoU thresholds (from 50\% to 95\%), providing a more comprehensive evaluation of detection performance. The YOLOv10 detector achieved consistently high detection performance across most surgical instrument classes, demonstrating robust localization under diverse laparoscopic conditions. The ROI class exhibited comparatively lower precision, recall, and mAP values because the detector was required to localize a dynamic operative workspace rather than a discrete object with well-defined boundaries. Variations in camera motion, zoom, tissue manipulation, smoke, specular reflections, and illumination lead to substantial changes in the ROI appearance and extent, increasing the difficulty of consistent bounding-box prediction. Moreover, because the ROI occupies a large portion of the image, small localization errors result in a greater reduction in IoU, particularly at higher evaluation thresholds, leading to a lower mAP$_{50-95}$ score. Nevertheless, the detector consistently localizes the operative workspace and provides meaningful spatial context for the subsequent stages of the proposed framework. An additional five-fold cross-validation study was performed using the data partitioning strategy described in Section II-A to further evaluate the robustness of the detector and the consistency of ROI-aware detection. Here, the model achieved a mean mAP@50 of 71.01 ± 2.62\% and a mean mAP@50--95 of 53.30 ± 2.20\% with the extended dataset (17 classes + ROI), demonstrating consistent detection performance across different data partitions and confirming the stability of the proposed detection framework (refer to Table \ref{tab:cv_results}).

\begin{table}[!t]
\caption{Five-fold cross-validation performance of the YOLOv10 detector on the hSDB dataset.}
\label{tab:cv_results}
\centering
\small
\setlength{\tabcolsep}{6pt}
\renewcommand{\arraystretch}{1.1}
\begin{tabular}{lcc}
\hline
\textbf{Training Dataset} & \textbf{mAP$_{50}$ (\%)} & \textbf{mAP$_{50-95}$ (\%)}\\
\hline
Original (17 classes) & $75.47 \pm 3.16$ & $57.31 \pm 2.12$ \\
Extended (17 classes + ROI) & $71.01 \pm 2.62$ & $53.30 \pm 2.20$ \\
\hline
\end{tabular}
\end{table}

\subsection{Tracking Performance}
\subsubsection{Quantitative Results}
Table \ref{tab2} compares the object detection performance of the chosen YOLOv10 detector with existing state-of-the-art models on the hSDB dataset\cite{yoon2021hsdb}. It achieved a modest yet consistent improvement over YOLOv5 and substantially outperformed Faster R-CNN, Cascade R-CNN, FCOS, and FoveaBox. Furthermore, incorporating the manually annotated ROI class significantly improved the detector's ability to localize the operative workspace, providing essential spatial context for subsequent occlusion analysis and depth-aware visualization. The best-performing detector was evaluated on the local dataset during inference, where it achieved an average IoU of 0.5570 and a mean confidence score of 0.8506, demonstrating reliable frame-to-frame localization and robust instrument tracking under realistic surgical conditions. The integration of particle filtering further improved temporal consistency during instrument motion and temporary occlusions, while the addition of calibrated depth estimation enhanced spatial interpretation of instrument–tissue interactions. Representative tracking results are illustrated in Fig. \ref{fig3}, showing the temporal progression of instrument localization at four representative time instances. A similar qualitative evaluation on the ovarian cystectomy video is provided in the supplementary material (Fig. S2). Supplementary videos further demonstrate the proposed framework by simultaneously visualizing surgical instrument detection, temporal tracking, occlusion events, and calibrated depth estimates during MIS procedures.

\begin{figure*}[!t]
\centerline{\includegraphics[width= 0.9 \textwidth]{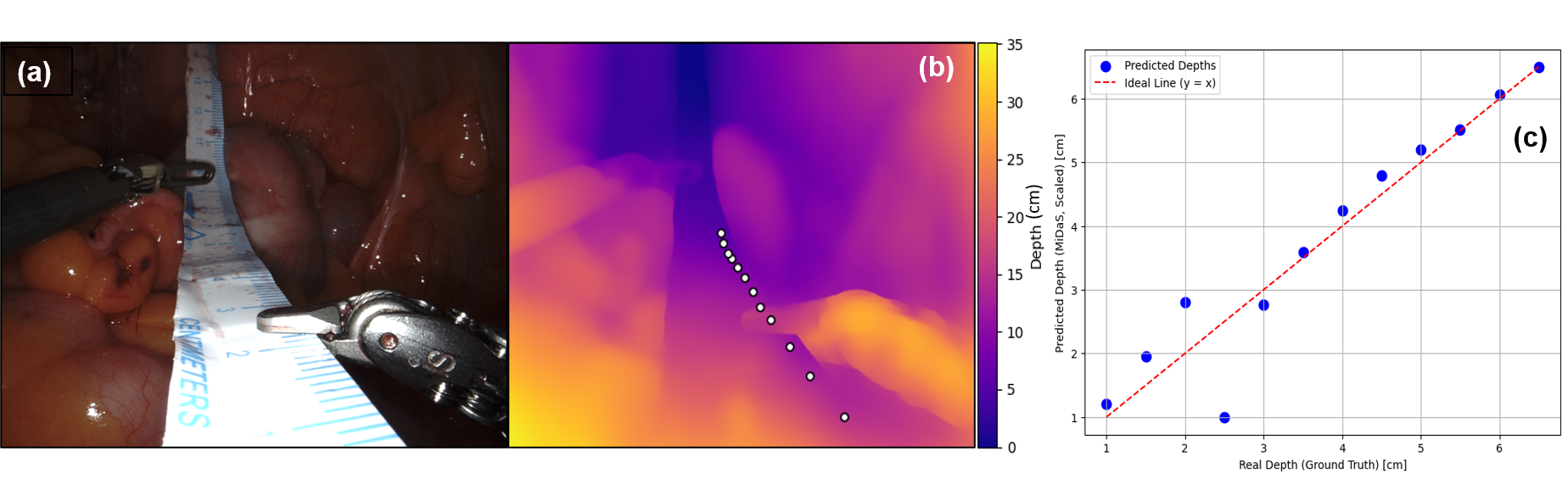}}
\caption{Depth mapping and validation during robotic surgery: (a) Reference surgical scene with paper scale, (b) Corresponding depth map with marked coordinates, (c) The linear regression plot of real and predicted depth values.}
\label{fig5}
\end{figure*}

\begin{table}[!t]
\caption{Comparison of object detection performance with SOTA methods on the hSDB dataset.}
\label{tab2}
\centering
\scriptsize
\setlength{\tabcolsep}{4pt}
\renewcommand{\arraystretch}{1.1}
\begin{tabular}{p{1.8cm} p{2.1cm} p{2.0cm} c}
\hline
\textbf{Method} & \textbf{Backbone} & \textbf{Configuration} & \textbf{mAP$_{50-95}$ (\%)}\\
\hline

Faster R-CNN & ResNet50 & FPN & 20.2\\
 & ResNet50 & FPN + OHEM & 20.1\\
 & ResNet50 & FPN + Double Head & 22.9\\
 & ResNet101 & FPN & 22.8\\
 & ResNeXt101-32$\times$4d & FPN & 24.1\\
 & ResNeXt101-64$\times$4d & FPN & 24.9\\
 & HRNetV2p-W18 & -- & 23.4\\
 & HRNetV2p-W40 & -- & 24.1\\
\hline

Cascade R-CNN & ResNet101 & FPN & 24.7\\
 & ResNeXt101-64$\times$4d & FPN & 24.5\\
 & HRNetV2p-W32 & -- & 25.7\\
\hline

FCOS & HRNetV2p-W32-GN & FPN & 11.8\\
 & ResNeXt101-64$\times$4d & FPN + MS Train & 22.0\\
\hline

FoveaBox & ResNet50 & FPN & 22.9\\
 & ResNeXt101 & FPN + Align-GN-MS & 23.7\\
\hline

YOLOv5 & -- & Without ROI & 28.7\\
 & -- & With ROI & 68.2\\
\hline

\textbf{YOLOv10} & \textbf{--} & \textbf{Without ROI} & \textbf{28.9}\\
 & \textbf{--} & \textbf{With ROI} & \textbf{71.3}\\
\hline

\end{tabular}

\vspace{1mm}
\end{table}

\subsubsection{Qualitative Comparison with Tracking and Occlusion Handling Methods}
The tracking performance was compared against representative tracking approaches while employing the same YOLOv10 detections for all methods. The evaluated methods include SORT \cite{bewley2016simple}, DeepSORT \cite{wojke2017simple}, ByteTrack \cite{zhang2022bytetrack}, and the proposed Particle Filter integrated with MiDaS depth estimation. Table~\ref{tab:tracking_comparison} summarizes the functional characteristics of these representative methods.

\begin{table*}[!t]
\caption{Qualitative comparison of representative tracking and occlusion handling approaches.}
\label{tab:tracking_comparison}
\centering
\setlength{\tabcolsep}{9pt}
\begin{tabular}{lcccccc}
\hline
\textbf{Method} &
\textbf{Motion Model} &
\textbf{Appearance} &
\textbf{Monocular} &
\textbf{Depth-aware} &
\textbf{Multi-object} &
\textbf{Real-time} \\
 &
 &
\textbf{Association} &
\textbf{Depth} &
\textbf{Occlusion Reasoning} &
\textbf{Tracking} &
\textbf{Suitable} \\
\hline

SORT \cite{bewley2016simple}
& Kalman Filter
& $\times$
& $\times$
& $\times$
& \checkmark
& \checkmark \\

DeepSORT \cite{wojke2017simple}
& Kalman + Deep Re-ID
& \checkmark
& $\times$
& $\times$
& \checkmark
& Limited \\

ByteTrack \cite{zhang2022bytetrack}
& IoU-based Association
& $\times$
& $\times$
& $\times$
& \checkmark
& \checkmark \\

\textbf{PF + MiDaS (Proposed)}
& \textbf{Particle Filter}
& $\times$
& \checkmark
& \checkmark
& \checkmark
& \checkmark \\

\hline
\end{tabular}
\end{table*}

\subsection{Ablation Study}
To isolate the specific contribution of the PF module from the overall framework, an ablation study was conducted using the local dataset. The best-performing \text{YOLOv10} detector weights, identical video sequences, and the same evaluation protocol were maintained across all experiments, with only the tracking module varied. Specifically, standalone \text{YOLOv10} detection was compared against the combined \text{YOLOv10 and PF} tracking framework. As summarized in Table~\ref{tab:ablation_results}, standalone \text{YOLOv10} achieved a Temporal Intersection over Union ($\mathrm{tIoU}$) of $0.4079$ with a mean detector confidence of $0.6680$. Since the detector processes each frame independently without temporal state propagation, transient detector failures caused by cautery smoke, specular reflections and partial instrument occlusions may result in bounding-box jitter and temporary loss of target localization. By incorporating the \text{YOLOv10 + PF} module, the $\mathrm{tIoU}$ increased to $0.6134$, while the mean detector confidence improved to $0.8339$. Using a weighted set of 100 particles to recursively estimate the object state, the PF maintains trajectory continuity during short-term detector dropouts and produces more stable spatial localization across consecutive frames.

\begin{table}[!t]
\caption{Ablation Study evaluating the contribution of Particle Filtering.}
\label{tab:ablation_results}
\centering
\footnotesize
\setlength{\tabcolsep}{5pt}
\renewcommand{\arraystretch}{1.1}
\begin{tabular}{lcccccc}
\hline
\textbf{Configuration} & \textbf{tIoU} & \textbf{Mean Confidence} & \textbf{Latency (ms)} & \textbf{FPS} \\
\hline
YOLOv10  & 0.4079 & 0.6680 & 18.3 & 54.6 \\
YOLOv10 + PF & \textbf{0.6134} & \textbf{0.8339} & \textbf{18.6} & 53.8 \\
\hline
\end{tabular}
\end{table}

\subsection{Occlusion Analysis}
The proposed occlusion module attained an overall frame-wise accuracy of 81.95\% with a mean detection delay of 16.6 ms which demonstrates the capability for near real-time identification of occlusion events. The evaluation was performed on 770 manually annotated frames extracted from the local dataset. Of these, only 28 frames corresponded to true occlusion events, resulting in a highly imbalanced dataset. Consequently, frame-wise precision and F1-score were adversely affected by the large proportion of non-occlusion frames. To minimize the likelihood of missed occlusion events, the framework was intentionally configured to operate with a predefined threshold \(\delta\) that prioritizes early occlusion detection, accepting a higher false-positive rate in exchange for improved sensitivity. This design choice is clinically motivated, as promptly identifying a visual obstruction is more desirable than missing an occlusion event that may temporarily compromise the surgeon's field of view.

Furthermore, to enhance intraoperative awareness and provide intuitive visual guidance, this module incorporates three visualizing mechanisms (Fig. \ref{fig3}). First, a text-based alert ("Warning: Occlusion Detected") is displayed immediately upon detecting an occlusion. Second, the instrument bounding box changes to red to provide a direct visual indication of occlusion. Third, depth-aware color-coded cues are employed, where the instrument bounding box is dynamically updated based on its proximity to the ROI: green indicates a safe region, yellow denotes partial occlusion, and red represents complete occlusion. A preliminary surgeon feedback survey has been conducted to assess the usability of these visualization mechanisms.

\subsection{Analytical Comparison of Monocular Depth Models}
To justify the selection of the monocular depth module, we performed an analytical comparison between the chosen MiDaS model and representative lightweight and transformer-based monocular depth estimation networks reported in the literature, namely FastDepth \cite{wofk2019fastdepth}, MonoDepth2 \cite{godard2019digging}, Depth Anything V2 Small \cite{yang2024depth}, and DPT-Hybrid \cite{ranftl2021vision}. The comparison considers model complexity, depth inference latency, estimated end-to-end pipeline latency within the framework, and the resulting throughput. For MiDaS, the parameter count and inference latency were obtained through local profiling on our experimental platform. For the remaining methods, model characteristics were obtained from the official model implementations. The end-to-end pipeline latency and throughput for the alternative models were analytically derived by replacing only the depth estimation module in our proposed framework while keeping the remaining components (YOLOv10 detection, particle filtering, and rendering) unchanged.

A comparative analysis of representative monocular depth estimation models is provided in Table t1 as supplementary material. Lightweight convolutional models such as FastDepth and MonoDepth2 offer lower computational cost and higher estimated throughput, whereas transformer-based models generally provide richer scene representations at the expense of increased computational complexity. Based on this analysis, MiDaS was selected as it provides a favorable balance between computational efficiency and model capacity. Local profiling showed a depth inference time of 21.8 ms, resulting in an estimated end-to-end pipeline latency of 25.9 ms and an effective throughput of 38.7 FPS, thereby satisfying the real-time requirements of MIS.
\subsection{Depth Estimation}
\begin{figure}[!t]
\centerline{\includegraphics[width=\columnwidth]{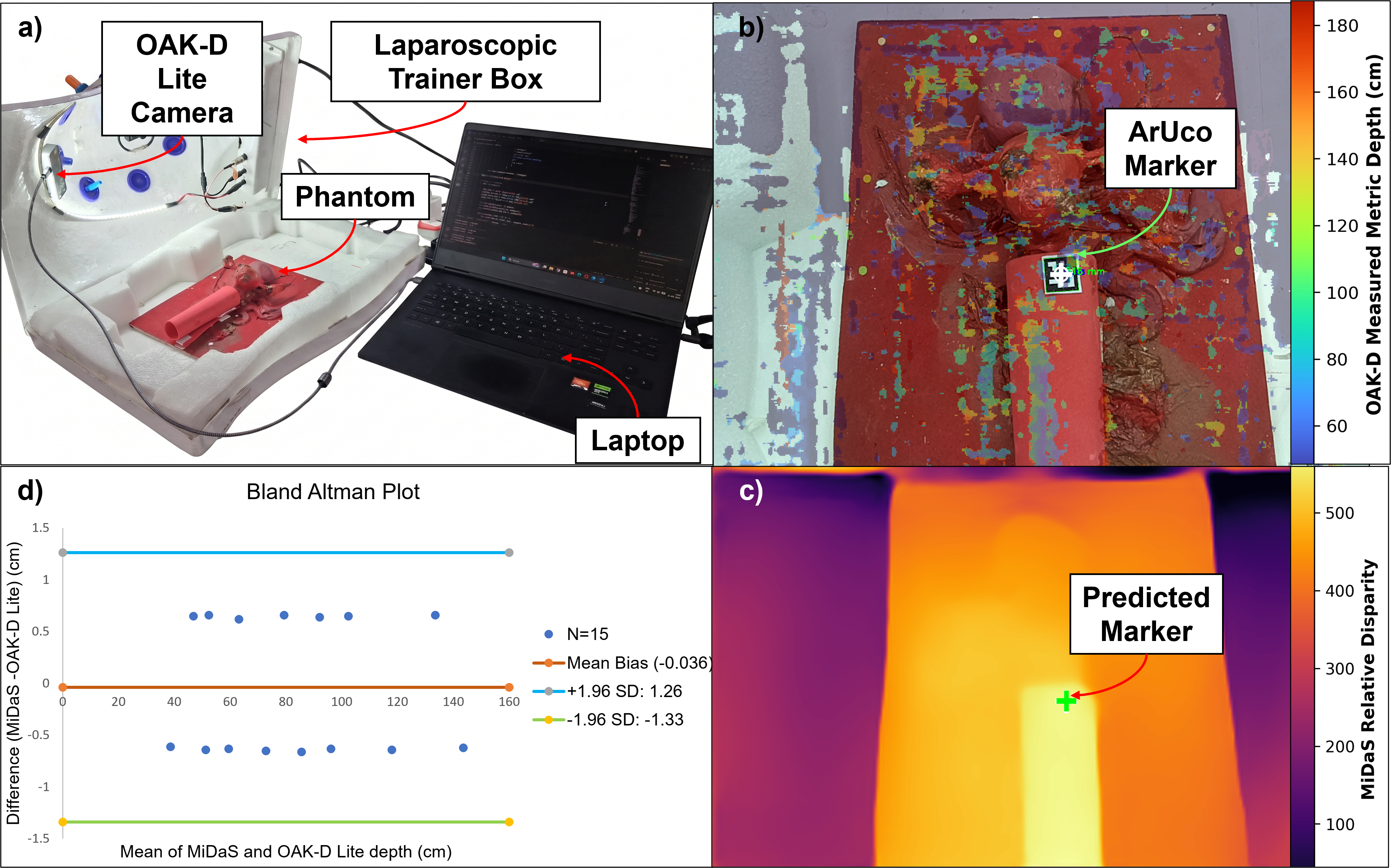}}
\caption{Independent phantom-based validation of monocular depth estimation using the OAK-D Lite camera (N = 15): (a) Experimental setup showing the phantom and imaging configuration, (b) Acquired depth map overlaid on the corresponding RGB image, with the automatically detected ArUco marker (51.6 cm), (c) Corresponding MiDaS-predicted depth map from the captured RGB image, yielding an estimated depth of 50.96 cm (absolute error = 0.64 cm), and (d) Bland–Altman agreement analysis between MiDaS-predicted depths and OAK-D Lite reference measurements across all validation samples.}
\label{fig8}
\end{figure}

The independent phantom-based experiment demonstrated good agreement between the calibrated MiDaS metric depth estimates and the reference stereo depth measurements acquired using the OAK-D Lite camera. Quantitative evaluation yielded an MAE of 0.65 cm and an RMSE of 0.66 cm, indicating that the proposed calibration framework provides reliable metric depth estimation across multiple phantom positions. The comparison between the calibrated MiDaS predictions and the reference stereo measurements is illustrated in Fig. \ref{fig8}(b)–(d), including the Bland–Altman analysis, which shows good agreement between the two measurement methods.

The primary limitation encountered during the validation was the reduced near-field performance of the stereo depth camera. As documented by the manufacturer, the OAK-D Lite is not optimized for close-range stereo depth sensing, making reliable depth estimation increasingly challenging at the short object-to-camera distances commonly encountered in MIS \cite{luxonis_oakdlite}\cite{banerjee2025surgiklar}. Consequently, occasional unstable or invalid stereo depth measurements were observed during the experiments. Despite this limitation, the calibrated MiDaS metric depth estimates remained in good agreement with the reference stereo measurements, providing additional confidence in the proposed depth calibration framework.

Following the independent validation, the calibrated MiDaS model was applied to the local surgical dataset for metric depth estimation. A sterile paper scale with known physical dimensions was introduced into the surgical field to establish the relationship between the relative MiDaS predictions and real-world distances. The calibration process achieved an MAE of 0.34 cm and an RMSE of 0.53 cm, indicating accurate depth estimation under clinical imaging conditions. The selected depth range of 1–6.5 cm represents the approximate working distance between the laparoscopic camera and the region of interest during minimally invasive surgery. The linear relationship between the measured and predicted depth values is illustrated in Fig. \ref{fig5}(c).

\subsection{System Latency}
\begin{figure}[!t]
\centerline{\includegraphics[width=\columnwidth]{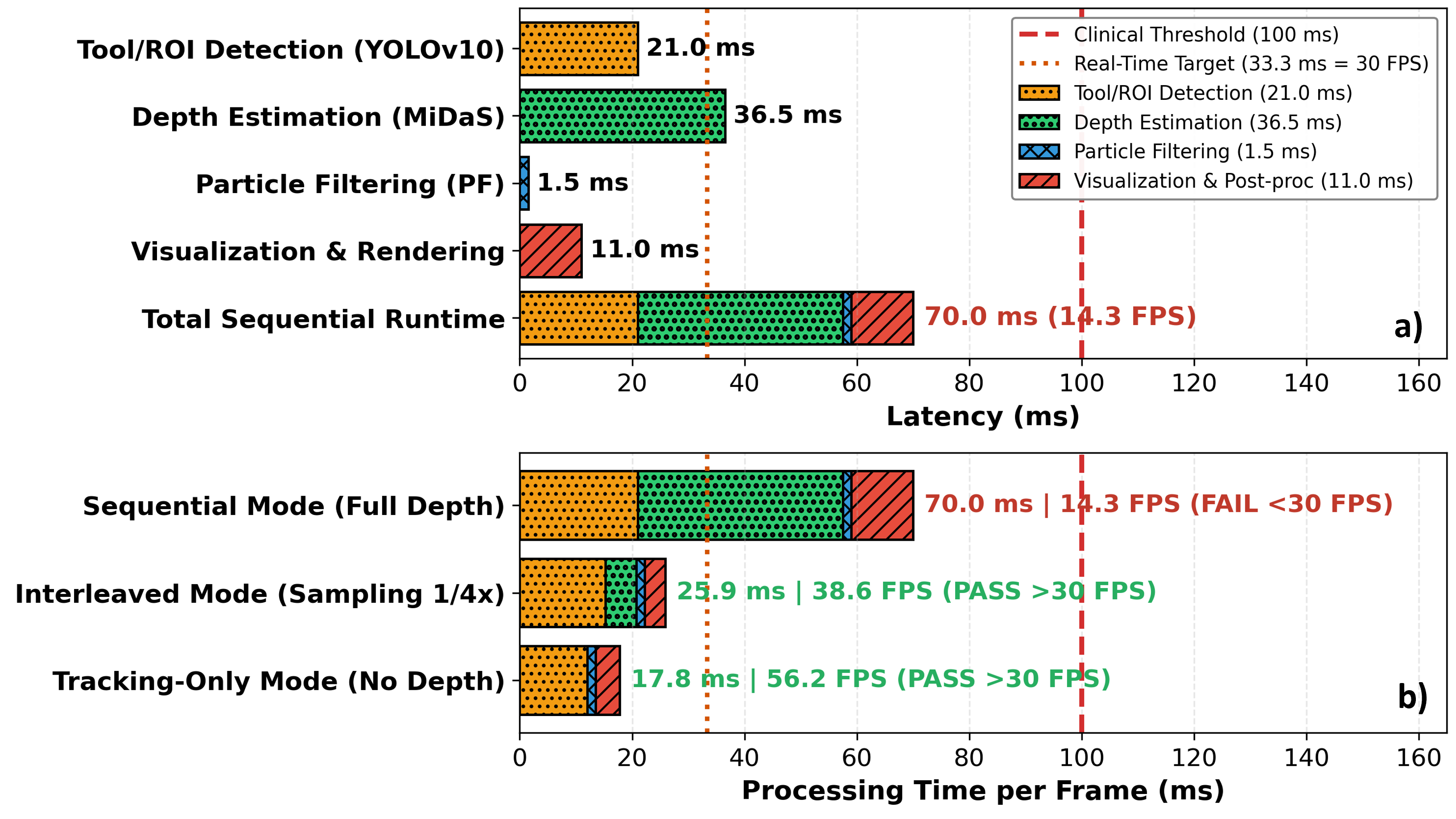}}
\caption{Runtime distribution and throughput comparison across pipeline operating modes: (a) Individual module latency breakdown for full sequential processing ($70.0$\,ms/frame); (b) Latency and frame rate comparison across Sequential ($14.3$\,FPS), Interleaved Depth Sampling ($38.7$\,FPS), and Tracking-Only ($56.1$\,FPS) operational modes against the $30$\,FPS real-time target ($33.3$\,ms) and $100$\,ms clinical threshold.}
\label{fig6}
\end{figure}

The complete processing pipeline executed over the local dataset is approximately 70 ms per frame under conventional sequential execution. Of this total latency, the combined neural network inference time, comprising YOLOv10 object detection and MiDaS model, accounted for approximately 34 ms per frame, while the remaining execution time was attributed to image preprocessing, PF tracking, occlusion analysis, visualization, and post-processing operations. A detailed breakdown of the runtime components revealed that MiDaS depth estimation constituted the largest computational overhead, followed by YOLOv10 object detection and particle filtering, with additional time attributed to preprocessing, occlusion detection, visualization, and post-processing tasks. The distribution is illustrated in Fig. \ref{fig6}, which shows the proportional contribution of each module to the overall pipeline latency. 

To achieve real-time performance, an Interleaved Depth Sampling strategy was adopted. Since anatomical background depth changes at a substantially lower temporal frequency than surgical instrument motion, MiDaS depth estimation was computed once every fourth frame, while YOLOv10 object detection and particle filter tracking continued to operate on every frame. This optimization reduced the effective per-frame depth computation from 36.5 ms to an average of 5.5 ms by executing MiDaS only once every four frames. The overall pipeline latency decreased to 25.9 ms per frame, resulting in an effective throughput of 38.7 FPS, thereby satisfying the real-time requirement of more than 30 FPS for intraoperative visualization. Furthermore, a tracking-only mode, in which depth estimation was omitted, achieved the lowest latency of 17.8 ms per frame, corresponding to 56.1 FPS. Despite the inclusion of multiple processing modules, the total end-to-end latency across all operating modes remained well below the clinically acceptable threshold of 100–160 ms, as documented in the literature for real-time laparoscopic video feedback \cite{heiliger2023ar,wild2020ieee}. The achieved runtime aligns closely with industry-standard surgical camera systems operating at 30–60 FPS, which tolerate processing delays of 50–100 ms without degrading surgical workflow \cite{celdran2025use}.

\subsection{Preliminary Usability Feedback}
To assess the clinical relevance and usability of the proposed framework, a structured multi-center feedback survey was conducted among experienced medical professionals, including specialty surgeons. A total of 13 respondents with varied surgical expertise including open surgery, laparoscopic, and robotic procedures, have participated in the survey. Majority had less than 5 years of experience, while 4 had more than 5 years. Most participants reported frequent challenges related to instrument-induced occlusions during laparoscopic surgeries and unanimously acknowledged that an automatic occlusion detection system would enhance both surgical safety and the decision-making process. They also highlighted its utility in intra-operative guidance, post-operative video review, and surgical training. Regarding the preference of occlusion alerts, majority had favored the color-coded bounding box, citing its intuitive visual cue and real-time usefulness. Many participants emphasized the potential value of integrating both occlusion and distance-based depth alerts for enhanced situational awareness. Additional feedback highlighted possible augmentations, such as guidance to achieve an occlusion-free view, text or voice-based prompts\cite{shen2024ringgesture}\cite{martin2023medical}.

\begin{figure}[!t]
\centerline{\includegraphics[width=\columnwidth]{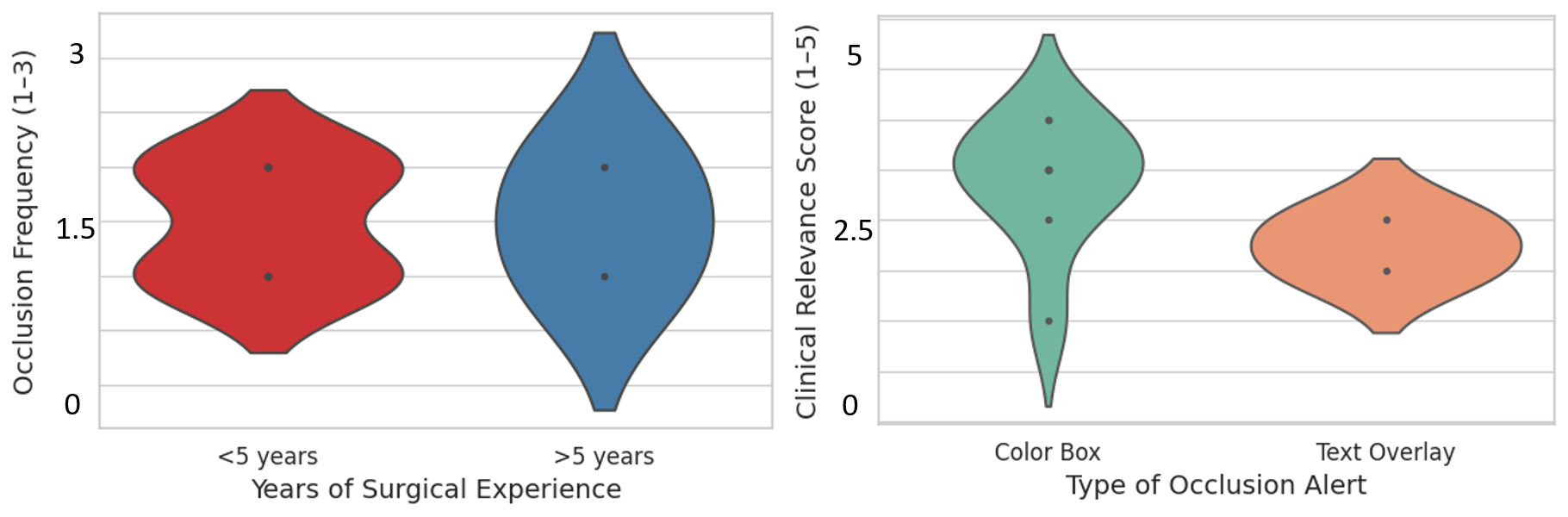}}
\caption{Violin plots of (a) Occlusion frequency and years of surgical experience, (b) Surgeons' preference of occlusion alerts.}
\label{fig7}
\end{figure}

Fig. \ref{fig7}(a) shows the violin plot comparing the frequency of occlusion-related challenges across surgical experience groups, measured on a scale of 1–3 (1 = low, 3 = high). Surgeons with less than 5 years of experience reported slightly higher frequencies, though distributions overlap across both groups. Fig. \ref{fig7}(b) illustrates the clinical relevance scores of different occlusion alerts on a scale of 1–5 (1 = least relevant, 5 = highly relevant). The color-coded bounding box shows a higher median and narrower spread, indicating it is perceived as more clinically intuitive and useful than text-based overlays, which received lower and more dispersed scores.

Although the present survey demonstrates encouraging clinical acceptance of the proposed framework, it does not establish whether the system improves objective surgical performance. The primary purpose of the survey was to assess the perceived clinical relevance, usability, and preferred alert mechanisms of the proposed depth-aware occlusion guidance system rather than to quantify improvements in operative outcomes. Heiliger et al. \cite{heiliger2023ar} evaluated augmented reality visualization techniques for laparoscopic guidance using an objective depth-sorting task in conjunction with subjective usability assessments, demonstrating that subjective user preference and objective task performance represent complementary but distinct aspects of clinical evaluation. Motivated by this approach, our future work will focus on integrating the proposed framework as a real-time augmented reality overlay on live laparoscopic video and evaluating its impact during standardized surgical skill exercises, such as the Fundamentals of Laparoscopic Surgery (FLS) peg-transfer task or equivalent box-trainer protocols. Quantitative performance measures, including task completion time, error rate, economy of motion, instrument path length, and occlusion-related events, together with validated workload and usability questionnaires, will be used to determine whether the proposed framework provides measurable improvements over conventional laparoscopic visualization.

\section{Discussion}
The proposed framework addresses a critical challenge in laparoscopic and MIS. The motivation behind this work is to overcome the limitations of monocular laparoscopic videos, which inherently lack 3D spatial cues. In visually constrained environments, both depth perception and occlusion handling become significant issues for surgeons and computational models alike. Occlusions arise when instruments obstruct the view of critical anatomical structures, complicating eye-hand coordination of surgeons. By incorporating monocular depth estimation into the tracking pipeline, the system can infer relative depth from a single RGB frame. This integration supports real-time, depth-aware feedback and visualization, while detecting tools during occlusion events, thereby enhancing surgical awareness and decision-making.

The fused architecture of deep learning based object detection and particle filtering ensures that tracking and visualization remain robust and continuous even in the presence of brief occlusions. Compared to existing approaches, our model demonstrates improved detection accuracy and temporal consistency, underscoring the importance of annotating ROIs for enhanced spatial understanding and occlusion handling in surgical scenes. Our system’s performance aligns well with industrial-grade surgical imaging systems, maintaining real-time responsiveness and operating within acceptable latency thresholds to ensure seamless intraoperative visual feedback and guidance. The preliminary clinical survey further supports the perceived clinical utility and relevance of the proposed framework, with participating surgeons recognising the potential value of real-time depth-aware occlusion alerts for intra-operative guidance, surgical training, and post-operative video analysis. Nevertheless, objective task-based validation using standardized laparoscopic skill assessments remains necessary to determine whether these perceived benefits translate into measurable improvements in surgical performance and workflow efficiency. 

In summary, the modular and adaptable nature of this framework makes it suitable for a variety of surgical environments and instrumentation setups, representing a meaningful step toward the development of intelligent assistive visualization technologies that support safer and more precise surgical interventions.

\section{Conclusion}
In this study, we presented a hybrid framework that combines particle filtering with real-time detection and monocular depth estimation to enhance occlusion handling and visualization in MIS. The approach was validated through comparative experiments and demonstrated superior performance over existing methods, achieving higher detection accuracy and enhanced temporal stability. The results emphasized the significance of dataset relevance and contextual quality for surgical instrument tracking and enhanced visualization. The depth estimation component also demonstrated satisfactory accuracy, confirming its effectiveness in estimating clinically meaningful depths. Beyond technical validation, we conducted a structured survey to evaluate the clinical relevance of the system. Survey participants affirmed the potential of this system to support intraoperative decision-making, reduce the risk of injury to critical structures, and aid in post-operative reviews and surgical training. These insights will guide the future development and refinement of the system, moving it closer to clinical deployment as an assistive visualization tool for safer and more informed surgical interventions.

\section{Author Contribution Statement}
AS and PA jointly conceptualized the study. AS handled the dataset, developed the algorithms for tool tracking, and depth estimation. PA developed the visualization techniques. AS, PA, VN contributed to the computational analysis. KB provided the clinical datasets and supported the clinical validation. SM supervised the project. All authors participated in preparation
of the manuscript.

\bibliographystyle{IEEEtran}
\bibliography{references}

\end{document}